\documentclass[a4paper,twoside]{article}

\usepackage{amsmath,amssymb}
\usepackage{microtype}
\usepackage{pslatex}
\usepackage{graphicx}
\usepackage{booktabs}
\usepackage{tikz}
\usetikzlibrary{arrows.meta,positioning,decorations.pathreplacing}
\usepackage{seqsplit}
\usepackage{soul}
\usepackage{needspace}
\usepackage[bottom]{footmisc}
\usepackage[hidelinks]{hyperref}
\usepackage[nameinlink,capitalise]{cleveref}
\usepackage{algorithm2e}
\usepackage{scitepress/apalike}
\usepackage{scitepress/SCITEPRESS}

\newif\ifblind
\blindfalse         % camera-ready / arXiv

\begin{document}

\title{Bit-Flip Vulnerability of Shared KV-Cache Blocks\\in LLM Serving Systems}

\ifblind
\author{\authorname{Anonymous Author(s)}
\affiliation{}
\email{}
}
\else
\author{\authorname{Yuji Yamamoto\sup{1} and Satoshi Matsuura\sup{1}}
\affiliation{\sup{1}Institute of Science Tokyo, Tokyo, Japan}
\email{yamamoto.y.dd03@m.isct.ac.jp, matsuura@cii.isct.ac.jp}
}
\fi


\abstract{Rowhammer on GPU DRAM has enabled adversarial bit flips in model
weights; shared KV-cache blocks in LLM serving systems present an
analogous but previously unexamined target.  In vLLM's Prefix Caching, these
blocks exist as a single physical copy without integrity protection.
Using software fault injection under ideal bit targeting, we
characterize worst-case severity and identify three properties:
(1)~Silent divergence---13 of 16 BF16 bit positions produce coherent
but altered outputs, indistinguishable from legitimate responses
without a clean baseline.
(2)~Selective propagation---only requests sharing the targeted prefix
are affected.
(3)~Persistent accumulation---no temporal decay occurs, so cumulative
damage grows linearly with subsequent requests.
Together, these constitute a threat profile distinct from weight
corruption: silent divergence and selective propagation enable
detection evasion; persistent accumulation then proceeds unchecked,
yielding damage amplification bounded only by how long the block
remains cached.
A checksum-based countermeasure detects any single-bit corruption at
scheduling time, bounding cumulative damage to one batch
independent of the block's cache lifetime, with negligible overhead.
These results argue for integrity protection of prefix blocks
before end-to-end exploitation is demonstrated.}

\onecolumn \maketitle \normalsize \setcounter{footnote}{0} \vfill

\section{\uppercase{Introduction}}
\label{sec:introduction}

As large language models (LLMs) are increasingly deployed as
production inference services, caching key--value (KV) tensors in GPU
memory---so that requests sharing a common prompt prefix need not
recompute them---has become standard practice.
Because these cached blocks exist as a single physical copy shared across
all referencing requests, one corruption event, such as a
hardware-induced bit flip, silently affects every request that reads
the block.  Yet no integrity mechanism (e.g., checksums or ECC) guards
this shared data.

The most widely deployed implementation is \emph{Prefix Caching} in
vLLM~\cite{kwon2023pagedattention}, the \emph{de facto} standard
for open-source LLM serving.  When many requests share an identical
prompt prefix (e.g., a system prompt or few-shot examples), vLLM
computes the corresponding KV blocks once and reuses them for all
subsequent requests.  We call these shared blocks
\emph{prefix blocks} (defined precisely in \cref{sec:bg-vllm}).
Enabled by default since vLLM v0.6.0, Prefix Caching makes prefix
blocks a pervasive component of production serving.

If a bit flip reaches a prefix block, Prefix Caching's design and the
structure of BF16 encoding give rise to three properties that
distinguish prefix-block corruption from weight corruption.
First, \emph{silent divergence}: the BF16
encoding~(\cref{sec:bg-bf16}) makes most bit positions produce only
small perturbations, so outputs remain coherent but differ from the
original---undetectable without a clean baseline.
Second, \emph{selective propagation}: corruption is confined to
requests sharing the targeted prefix; co-located workloads remain
unaffected.
Third, \emph{persistent accumulation}: corrupted blocks are never
invalidated during normal operation, so cumulative damage grows
linearly with subsequent requests throughout the block's
\emph{residency} (the duration it remains cached before eviction).
Together, these properties constitute a threat profile qualitatively
distinct from that of weight corruption.

Prior studies~\cite{alnahian2025cachetrap,ganesh2025kvcachemanipulation,wu2026cacheme,xia2025kelle}
have neither systematically characterized the severity of shared
prefix-block corruption nor quantified how these three properties
compound into a distinct threat profile (\cref{sec:related-work}).
Meanwhile, recent work has narrowed---but not closed---the gap
between demonstrated bit-flip primitives and end-to-end
exploitation~(\cref{sec:bg-rowhammer,sec:discussion-limitations}).

To characterize worst-case severity before this gap is closed,
we use software fault injection (SFI) under an ideal-bit-targeting
assumption (\cref{sec:threat-model}).
We additionally propose a lightweight countermeasure that bounds
cumulative damage.
Our contributions are:
\begin{enumerate}
  \item \textbf{Bit-position sensitivity map.}  A systematic scan of
    all 16 bit positions of BF16 (2{,}400 trials, Qwen3-8B) shows that
    13 of 16 positions yield silent divergence\footnote{Bits~0--11
    and~15, all with collapse rates below 1\%;
    see \cref{sec:eval-sensitivity}.}, with collapse concentrated
    in the upper exponent bits.

  \item \textbf{Selective propagation verification.}  A mixed-prefix
    experiment (120 trials) confirms that corruption affects only
    requests referencing the corrupted block; co-located requests
    with a different prefix are entirely unaffected.

  \item \textbf{Temporal persistence analysis.}  Corruption tracking
    over 24{,}000 post-injection requests across two models shows no
    temporal decay; cumulative damage grows strictly linearly with
    subsequent requests, bounded only by the block's residency.

  \item \textbf{Checksum-based countermeasure.}  A lightweight
    scheduling-time integrity check that detects any single-bit
    corruption, bounding cumulative damage to one batch independent
    of block residency, with negligible throughput overhead.
\end{enumerate}
The remainder of this paper is organized as follows.
\Cref{sec:background,sec:threat-model,sec:methodology} cover
background, the threat model, and our SFI methodology;
\cref{sec:evaluation,sec:countermeasures} present the severity
characterization and the checksum countermeasure; and
\cref{sec:discussion,sec:related-work,sec:conclusion} cover
discussion, related work, and conclusions.

\section{\uppercase{Background}}
\label{sec:background}

\subsection{vLLM and Prefix Caching}
\label{sec:bg-vllm}

vLLM~\cite{kwon2023pagedattention} manages KV-cache memory in
fixed-size \emph{KV blocks}, each holding a predetermined number of
tokens (typically 16) and storing the per-token key and value vectors
across all attention heads.  In vLLM's production engine (v1), a fully
computed block is \emph{immutable by convention}---new tokens go to
freshly allocated blocks, while existing blocks are read but never
overwritten by the framework.  No hardware or OS mechanism enforces
this; the GPU memory remains fully writable, leaving cached blocks
vulnerable to external bit-flip faults.

\subsubsection{Block Management and GPU Layout}
On the CPU side, a \emph{block pool} manages block metadata,
including a hash-to-block map for prefix cache lookups and an LRU
free-block queue that governs eviction and reuse.
On the GPU side, KV-cache tensors are stored in a single contiguous
buffer of shape
$[2 \times N_{\text{blocks}} \times B \times H \times D]$, where
the leading dimension indexes keys ($k{=}0$) and values ($k{=}1$),
$B$ denotes the block size, $H$ the number of KV heads, and $D$ the
head dimension.  Block identifiers are fixed at initialization, so
each block's GPU virtual address is deterministic.

\subsubsection{Prefix Caching}
Prefix Caching leverages the fact that many requests to an LLM
service share a common prompt prefix---e.g., a system prompt or
few-shot examples.  When enabled (the default since vLLM v0.6.0),
the framework caches fully computed KV blocks and
indexes them by a \emph{chain hash}.  Each block's hash is computed
over the parent block's hash, the block's token IDs, and optional
extra keys (LoRA identifiers, multimodal input hashes, or a
per-request \texttt{cache\_salt}).  This chaining ensures that a
cache hit requires an exact token-by-token match of the entire
prefix up to and including the candidate block.

On each request, the scheduler looks up block hashes in order; a hit
\emph{touches} the existing physical block (incrementing its reference
count \texttt{ref\_cnt}), and the first miss triggers fresh allocation
for the remaining blocks.  While $\mathtt{ref\_cnt} > 0$, a block is
immune to eviction.  At completion, \texttt{ref\_cnt} is decremented;
one reaching zero retains its hash and stays available for future hits
until evicted by LRU\@.  Popular prefix blocks thus remain cached
indefinitely.

Not every cached block is reliably shared, however: a block at the
prefix--suffix boundary may incorporate request-specific tokens,
preventing cross-request reuse.  We therefore define
\emph{prefix blocks} as the full KV blocks that are reliably shared
across requests via Prefix Caching, excluding such boundary blocks.

\subsection{Bit-Flip Threats to GPU Memory}
\label{sec:bg-rowhammer}

Rowhammer~\cite{kim2014rowhammer} is a DRAM vulnerability in which
rapidly activating a row induces bit flips in physically adjacent
rows.  First demonstrated on commodity DDR3, it has since been
extended to a wide range of targets~(\cref{sec:related-work}).
GPU GDDR6 memory is no exception: recent work confirms
bit flips on GPUs~\cite{lin2025gpuhammer,coalson2025prisonbreak,lin2026gpubreach}.
Hardware ECC corrects single-bit and detects double-bit errors but is
defeated by three or more flips in a code word; moreover, GDDR-based
GPUs commonly leave ECC disabled by default and let tenants turn it
off~\cite{lin2025gpuhammer,lin2026gpubreach}, leaving GPU-resident
data, including KV-cache tensors, exposed to bit-flip corruption.

\subsection{BF16 Floating-Point Representation}
\label{sec:bg-bf16}

The bfloat16 (BF16) format~\cite{kalamkar2019bfloat16} is the
standard storage format for KV-cache tensors in current LLM serving
systems.  It uses 16~bits: a 1-bit sign ($S$), an 8-bit exponent
($E_7$--$E_0$), and a 7-bit mantissa ($M_6$--$M_0$), retaining the
full FP32 dynamic range while halving the memory footprint.
\Cref{tab:bf16-bits} maps each bit index to its semantic role.
This encoding creates a strong hierarchy in bit-flip impact,
with expected perturbation magnitude following the ordering
mantissa $<$ sign $<$ exponent.
A sign-bit flip yields a perturbation of $2|v|$---larger than any
mantissa shift, but proportional to the original value.
In the extreme case, an upper-exponent flip can push the biased
exponent to 255, producing \texttt{NaN} or $\pm\infty$.

\begin{table*}[t]
  \centering
  \caption{BF16 bit layout.  Each of the 16 bits is shown with its
    field (sign, exponent, or mantissa), symbol, and value-level
    weight; bit~0 is the LSB.}
  \label{tab:bf16-bits}
  \footnotesize
  \setlength{\tabcolsep}{3pt}
  \begin{tabular}{@{}l*{16}{c}@{}}
    \toprule
    Bit    & 15 & 14 & 13 & 12 & 11 & 10 & 9 & 8 & 7 & 6 & 5 & 4 & 3 & 2 & 1 & 0 \\
    \midrule
    Field  & \multicolumn{1}{c}{S}
           & \multicolumn{8}{c}{Exponent}
           & \multicolumn{7}{c}{Mantissa} \\
    \cmidrule(lr){3-10}\cmidrule(lr){11-17}
    Symbol & $s$ & $E_7$ & $E_6$ & $E_5$ & $E_4$ & $E_3$ & $E_2$ & $E_1$ & $E_0$
           & $M_6$ & $M_5$ & $M_4$ & $M_3$ & $M_2$ & $M_1$ & $M_0$ \\
    Weight & $\pm$ & $2^{7}$ & $2^{6}$ & $2^{5}$ & $2^{4}$ & $2^{3}$ & $2^{2}$ & $2^{1}$ & $2^{0}$
           & $2^{-1}$ & $2^{-2}$ & $2^{-3}$ & $2^{-4}$ & $2^{-5}$ & $2^{-6}$ & $2^{-7}$ \\
    \bottomrule
  \end{tabular}
\end{table*}

\section{\uppercase{Threat Model}}
\label{sec:threat-model}

We consider a multi-tenant GPU serving environment in which an LLM
inference provider deploys vLLM with Prefix Caching enabled on a shared
GPU\@.  Multiple tenants submit requests that share a common system
prompt, causing vLLM to cache the corresponding KV blocks and serve
them to all subsequent requests referencing the same prefix.

\subsection{Attacker Model}
The adversary is a co-located tenant whose goal is to corrupt prefix
blocks so that other tenants' responses silently deviate from expected
behavior---an integrity violation.
The adversary induces bit-flip corruption in GPU DRAM via hardware
fault injection, with Rowhammer~\cite{lin2025gpuhammer} as the primary
demonstrated mechanism (\cref{sec:bg-rowhammer}).  This requires no
software privileges beyond a normal tenant's and leaves model weights,
application code, and API-level inputs unaffected.

\subsection{Attack Surface}
The target is the value tensors of prefix blocks in vLLM's KV-cache.
Because a prefix block is a single physical copy shared across all
referencing requests without integrity protection, one bit flip
immediately affects all concurrent and future requests referencing it.
By default (\texttt{cache\_salt=None}) this sharing spans all tenants,
so the attacker and victim belong to the same salt group; vLLM's opt-in
salting~\cite{vllm_pr17045} narrows it only to organization or
tenant-group granularity, so co-tenants still share blocks under both
configurations.

\subsection{Scope}
We frame the vulnerability as an integrity violation, in which
undetected modification of KV-cache data causes outputs to deviate
from expected behavior.
A single BF16 bit flip can range from a sub-1\% mantissa perturbation
to an exponent overflow producing \texttt{NaN} or $\pm\infty$
(\cref{sec:bg-bf16}).  The key distinction is therefore between
\emph{collapse} (degenerate, overtly detectable output) and
\emph{silent divergence} (coherent but altered output, undetectable
without a clean baseline).

Physically reaching a specific prefix block via Rowhammer remains
undemonstrated.  We therefore characterize worst-case severity across
this spectrum under an \emph{ideal bit targeting} assumption (a single
bit at any position can be flipped), which deliberately relaxes
physical Rowhammer constraints; we do not claim or demonstrate
end-to-end exploitation.  The remaining gap is discussed in
\cref{sec:discussion-limitations}.
Confidentiality threats (e.g., prompt
extraction~\cite{wu2025promptpeek}) are outside our scope.

\section{\uppercase{Methodology}}
\label{sec:methodology}

\subsection{Software Fault Injection Framework}
\label{sec:method-sfi}

We characterize the KV-cache vulnerability through SFI under the
ideal-bit-targeting assumption stated in \cref{sec:threat-model}---the
standard methodology for characterizing vulnerability severity prior to
end-to-end
exploitation~\cite{rakin2019bfa,alnahian2025cachetrap}.
Our framework flips individual bits in the
BF16 representation of KV-cache tensors via XOR masking.

\subsubsection{Injection Target Selection}
We target the value tensors of prefix blocks, the most direct
propagation pathway: a value-side perturbation reaches the attention
output directly, whereas a key-side one is redistributed through
softmax normalization---consistent with
\textsc{CacheTrap}~\cite{alnahian2025cachetrap}.

The targeted prefix blocks are determined by the prompt
construction (\cref{sec:method-setup}).  Preliminary experiments
sampling all cached blocks confirmed that injections into prefix
blocks produced substantially higher output change rates
(${\sim}$64\%) than non-prefix blocks (${\sim}$7\%), as expected from
the shared single-copy design; we therefore restrict the injection
surface to prefix blocks.

\subsubsection{Injection Mechanism}
We instrument vLLM by dynamically patching the \texttt{execute\_model()}
method of \texttt{GPUModelRunner}.  At injection, the framework locates
the target KV-cache tensor---a contiguous GPU buffer of shape
$[2 \times N_\text{blocks} \times B \times H \times D]$---and
reinterprets the target element as a 16-bit integer.  A single-bit XOR
mask is then applied in-place:
\begin{equation}
  \label{eq:xor-flip}
  \hat{v} \;=\; v \oplus (1 \ll p),
\end{equation}
where $v$ is the original BF16 value (reinterpreted as
\texttt{int16}), $p \in \{0,\ldots,15\}$ is the target bit position,
and $\hat{v}$ is the corrupted value.  This modifies the shared
physical block directly, so all requests referencing it observe the
corrupted data on their next read---a direct consequence of the
single-copy, no-integrity-check design.  The resulting perturbations
are identical to those from Rowhammer-induced charge
leakage~\cite{kim2014rowhammer}.

\subsubsection{Trial Flow}
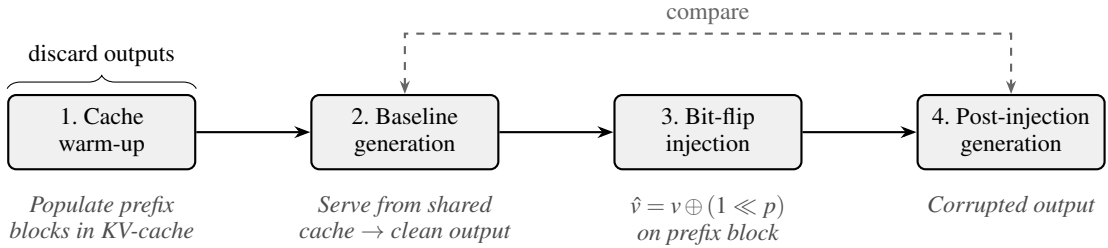
\begin{figure*}[t]
  \centering
  \begin{tikzpicture}[
    phase/.style={
      rectangle, rounded corners=3pt, draw=black, thick,
      minimum height=2.8em, minimum width=7em,
      font=\footnotesize, align=center, fill=black!6},
    arr/.style={-{Stealth[length=6pt]}, thick},
    note/.style={font=\footnotesize\itshape, text=black!70, align=center},
  ]
    % Phase nodes
    \node[phase] (warm) {1.~Cache\\warm-up};
    \node[phase, right=1.5cm of warm] (base) {2.~Baseline\\generation};
    \node[phase, right=1.5cm of base] (flip) {3.~Bit-flip\\injection};
    \node[phase, right=1.5cm of flip] (pois) {4.~Post-injection\\generation};

    % Arrows
    \draw[arr] (warm) -- (base);
    \draw[arr] (base) -- (flip);
    \draw[arr] (flip) -- (pois);

    % Annotations below
    \node[note, below=0.6em of warm]
      {Populate prefix\\blocks in KV-cache};
    \node[note, below=0.6em of base]
      {Serve from shared\\cache $\to$ clean output};
    \node[note, below=0.6em of flip]
      {$\hat{v} = v \oplus (1 \ll p)$\\on prefix block};
    \node[note, below=0.6em of pois]
      {Corrupted output};

    % Brace over phase 1 (outputs discarded)
    \draw[decorate, decoration={brace, amplitude=5pt, raise=2pt}]
      (warm.north west) -- (warm.north east)
      node[midway, above=7pt, font=\footnotesize] {discard outputs};

    % Compare arc from phase 2 to phase 4
    \draw[dashed, thick, {Stealth[length=5pt]}-{Stealth[length=5pt]}, black!60]
      (base.north) -- ++(0,0.8) -| (pois.north)
      node[pos=0.25, above, font=\footnotesize] {compare};
  \end{tikzpicture}
  \caption{Trial flow for a single experimental condition.  Each trial
    proceeds through four sequential phases; the warm-up outputs are
    discarded and only phases~2 and~4 contribute to the evaluation
    metrics.}
  \label{fig:trial-flow}
\end{figure*}

Each trial proceeds in four phases (\cref{fig:trial-flow}): warm-up,
baseline generation on the same requests, bit-flip injection, and
post-injection generation.  Injection targets bit position~$p$ while
the remaining coordinates (block, layer, head, token position,
channel) are randomized uniformly within the prefix blocks, isolating
the effect of $p$.  Post-injection outputs are then compared
token-by-token against the baseline
(\cref{sec:method-metrics}).

\subsubsection{Framework Validation}
We first establish the framework's noise floor: across 60
injection-free trials over six batch sizes
($n_c \in \{1,2,4,8,16,32\}$, 10 each), every output matched its
baseline token-by-token under deterministic decoding.  Any divergence
in the experiments is therefore attributable solely to injected bit
flips.

\subsection{Evaluation Metrics}
\label{sec:method-metrics}

We define three corruption-specific metrics and two standard
similarity metrics.

\subsubsection{Token Change Rate (TCR)}
TCR measures \emph{propagation breadth}: the fraction of $n_c$
concurrent requests whose token sequences change after injection.
\begin{equation}
  \label{eq:tcr}
  \mathrm{TCR} = \frac{1}{n_c}
    \sum_{i=1}^{n_c} \mathbb{1}\!\bigl[
      \operatorname{tok}(y_i) \neq \operatorname{tok}(\hat{y}_i)
    \bigr].
\end{equation}

\subsubsection{Token Diff Ratio (TDR)}
TDR measures per-request \emph{impact depth}: the fraction of token
positions that differ between clean and post-injection outputs,
with out-of-range positions treated as mismatches.
\begin{equation}
  \label{eq:tdr}
  \mathrm{TDR}_i = \frac{1}{L}
    \sum_{j=1}^{L} \mathbb{1}\!\bigl[
      t_j \neq \hat{t}_j
    \bigr],
\end{equation}
where $L = \max(|\operatorname{tok}(y_i)|,\,
|\operatorname{tok}(\hat{y}_i)|)$.

\subsubsection{Output Change Rate (OCR)}
OCR aggregates across $T$ trials under the same condition:
\begin{equation}
  \label{eq:ocr}
  \mathrm{OCR} = \frac{1}{T}
    \sum_{t=1}^{T} \mathbb{1}\!\bigl[
      \mathrm{TCR}_t > 0
    \bigr].
\end{equation}

\subsubsection{Similarity Metrics}
We report ROUGE-L~(F1 with stemming)~\cite{lin2004rouge} for
surface-level textual similarity and
BERTScore~(F1)~\cite{zhang2020bertscore} for semantic similarity.
ROUGE-L, a longest-common-subsequence (LCS) measure, is sensitive to
the localized substitutions,
insertions, and deletions characteristic of silent divergence, whereas
BERTScore captures the meaning-preserving paraphrases that ROUGE-L
underweights.
We do not adopt BLEU: designed for multi-reference machine
translation, it over-penalizes the legitimate reorderings of our
single-reference, long-form outputs.

\subsection{Experimental Setup}
\label{sec:method-setup}

We conduct three experiments, each targeting a different property of prefix-block corruption.

\subsubsection{Experimental Conditions}
\Cref{tab:experiments} summarizes the three experiments.
\begin{table}[t]
  \centering
  \caption{Summary of experiments.  $p$: target bit position;
    $n_c$: concurrent requests per trial;
    Trials: number of trials (30 per condition).}
  \label{tab:experiments}
  \small
  \resizebox{\columnwidth}{!}{%
  \begin{tabular}{@{}llr@{}}
    \toprule
    Experiment & Variable & Trials \\
    \midrule
    Bit-position sensitivity & $p$: 0--15, $n_c$: 2--32 & 2{,}400 \\
    Selective propagation & $p \in \{0,6,14,15\}$ & 120 \\
    Temporal persistence   & $p \in \{0,6,14,15\}$, 2 models & 240 \\
    \bottomrule
  \end{tabular}}
\end{table}
\textbf{Bit-position sensitivity} maps per-bit severity across the full
BF16 layout.
The remaining two target four representative positions spanning the
BF16 bit-significance hierarchy---mantissa LSB, mantissa MSB,
exponent MSB, and sign (\cref{sec:bg-bf16}).
\textbf{Selective propagation} compares two request groups referencing
different prefix sets, testing whether corruption is confined to the
block that is actually read.
\textbf{Temporal persistence} tracks corruption over 100 successive
requests on both models to test for decay.

\subsubsection{Hardware and Software}
All experiments run on an NVIDIA DGX Spark workstation equipped with
128\,GB of unified GPU memory.  We use vLLM
v0.18.1rc0\footnote{Commit \texttt{298e510};
the most recent version available at the time of the experiments.} with Prefix Caching
enabled (\texttt{enable\_prefix\_caching=True}) and cache salting
disabled (\texttt{cache\_salt=None}), maximizing cross-request block
sharing.  Decoding is deterministic (\texttt{temperature=0}) with a
maximum generation length of 128 tokens.

\subsubsection{Models}
We evaluate two open-weight 8B-parameter models stored in BF16:
Qwen3-8B~\cite{qwen2025qwen3} and
DeepSeek-R1-Distill-Llama-8B~\cite{deepseek2025r1}
(hereafter \emph{DeepSeek-R1}).
Qwen3-8B serves as the primary model; the exhaustive 16-position scan
(\cref{sec:eval-sensitivity}) uses it alone, since the per-bit
sensitivity hierarchy is expected to follow from the BF16 encoding
(\cref{sec:bg-bf16}) rather than model-specific properties.
DeepSeek-R1 is added in the temporal-persistence experiment
(\cref{sec:eval-persistence}) to check cross-model consistency.

\subsubsection{Prompt Construction}
Every request shares a common system-prompt prefix of 337 characters
(${\approx}$103~tokens), spanning seven contiguous KV blocks at block
size 16.  The first six blocks hold exactly 16 prefix tokens each and
are always eligible for cross-request cache hits; the seventh is a
boundary block that is not reliably shared, so the injection surface
comprises the first six blocks.  Per-request suffixes are drawn from
the ShareGPT conversational dataset~\cite{anon2023sharegpt}, providing
diverse user queries.

\section{\uppercase{Evaluation}}
\label{sec:evaluation}

\subsection{Bit-Position Sensitivity Analysis}
\label{sec:eval-sensitivity}

We systematically evaluate all 16 bit positions of BF16 across five
concurrency levels ($n_c \in \{2, 4, 8, 16, 32\}$) on Qwen3-8B,
yielding $16 \times 5 = 80$ conditions with 30 trials each (2{,}400
trials total).  In each trial, a single bit at position~$p$ is flipped
in the value tensor of a uniformly random element within the prefix
blocks.

\begin{figure}[t]
  \centering
  \includegraphics[width=\columnwidth]{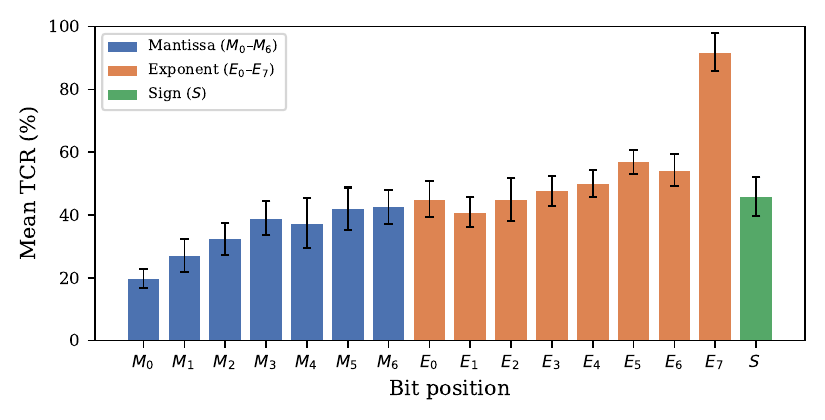}
  \caption{Mean TCR by bit position, averaged across all five
    concurrency levels (error bars: s.d.\ across $n_c$ levels).}
  \label{fig:tcr-by-bit}
\end{figure}

\subsubsection{Overall Sensitivity Hierarchy}
\Cref{fig:tcr-by-bit} shows that mean
TCR generally increases from mantissa LSB to exponent MSB, with the
sign bit falling between them---consistent with the BF16 perturbation
magnitude hierarchy (\cref{sec:bg-bf16}).
The remaining metrics in \cref{tab:bit-sensitivity} confirm this
ordering: TDR tracks the same hierarchy while ROUGE-L and BERTScore
decrease accordingly.  Except at bit~14, BERTScore
$\geq 0.93$ and ROUGE-L $\geq 0.71$ at every position, confirming
the silent divergence property hypothesized in
\cref{sec:threat-model}.

\subsubsection{Four-way Trial Classification}
\Cref{sec:threat-model} hypothesized two corruption modes---collapse
and silent divergence.  Based on TCR and mean ROUGE-L, we refine this
into four categories:
\begin{itemize}
  \item \textbf{No effect} ($\mathrm{TCR}=0$).
  \item \textbf{Partial divergence} ($0<\mathrm{TCR}<1$): some but not
    all concurrent requests affected.
  \item \textbf{Complete divergence} ($\mathrm{TCR}=1$,
    $\overline{\mathrm{ROUGE\text{-}L}}\geq 0.1$): coherent but altered
    natural language.
  \item \textbf{Collapse} ($\mathrm{TCR}=1$,
    $\overline{\mathrm{ROUGE\text{-}L}}<0.1$): incoherent or repetitive
    output.
\end{itemize}
The 0.1 threshold is chosen because collapse trials cluster near
$\mathrm{ROUGE\text{-}L}=0.0$, well separated from the lowest-scoring
divergence trials.
\Cref{tab:bit-sensitivity} reports the per-bit classification counts;
partial divergence is the dominant outcome across all 2{,}400 trials.

Collapse is overwhelmingly concentrated at bit~14 (${\sim}$72\% of all
collapse events); the remainder occurs exclusively in the upper
exponent bits (bits 11--13; only 1 event at bit 11), with none at any
mantissa or sign bit.  This is consistent with the BF16 encoding:
bit~14 flips can produce \texttt{NaN} or $\pm\infty$, which propagate
through attention as degenerate values, while lower exponent bits
typically yield perturbations within the representable range.

Complete divergence is rare and broadly distributed across exponent
and sign bits, representing cases where
the perturbation was large enough to shift generation to an
entirely different but linguistically valid trajectory.

\subsubsection{Qualitative Output Analysis}
The two dominant regimes produce qualitatively distinct outputs
(\cref{tab:output-examples}).  In the collapse regime (bit~14),
outputs degenerate into repetitive token sequences bearing no semantic
relation to the prompt, readily detectable by simple heuristics (e.g.,
repetition ratio or perplexity thresholds).  In the silent divergence
regime (bits~0--11 and~15), outputs are grammatically correct,
topically relevant, and often differ from the baseline only in minor
lexical or syntactic variations, making them indistinguishable from
legitimate responses; security implications are discussed in
\cref{sec:eval-interpretation}.

\begin{table}[t]
  \centering
  \caption{Representative corrupted outputs.  Underlined spans
    highlight differences from the baseline.}
  \label{tab:output-examples}
  \small
  \begin{tabular}{@{}p{0.95\columnwidth}@{}}
    \toprule
    \textbf{Silent divergence} (bit~3, $M_3$; $n_c{=}2$,
    $\mathrm{TCR}{=}1.0$, $\mathrm{ROUGE\text{-}L}{=}0.951$) \\
    \midrule
    \textbf{Baseline:} \textit{``\ldots The atom should have a name,
    such as ``Hydrogen'' or \ul{``Helium''}, displayed at the
    top.\ \ldots''} \\[4pt]
    \textbf{Corrupted:} \textit{``\ldots The atom should have a name,
    such as ``Hydrogen'' or \ul{``Oxygen''}, displayed at the
    top.\ \ldots''} \\
    \midrule
    \textbf{Collapse} (bit~14, $E_7$; $n_c{=}2$,
    $\mathrm{TCR}{=}1.0$, $\mathrm{ROUGE\text{-}L}{=}0.000$) \\
    \midrule
   \textbf{Baseline:} \textit{``How is it calculated?\ What are its
    advantages and disadvantages compared to impact factor?\ The
    Citation Success Index (CSI) is a metric designed to compare the
    citation capacity of pairs of journals\ldots''} \\[4pt]
    \textbf{Corrupted:}
    \texttt{\seqsplit{!!!!!!!!!!!!!!!!!!!!!!!!!!!!!!!!!!!!!!!!!!!!!!!!!!!!!!!!!!!!!!!!!!!!!!!!!!!!!!!!!!!!!!!!!!!!!!!!!!!!!!!!!!!!!!!!!!!!!!!!!!!!!!!!}} \\
    \bottomrule
  \end{tabular}
\end{table}

\subsubsection{Concurrency Independence}
The preceding analyses pool results across five concurrency levels;
to justify this, we test whether TCR depends on~$n_c$.
Mean TCR remains within a narrow range (43.9--46.7\%) across
$n_c \in \{2, 4, 8, 16, 32\}$, and Kruskal--Wallis tests show no
significant effect at either the aggregate ($H = 8.04$, $p = 0.090$)
or per-bit level after Bonferroni correction (\cref{tab:bit-sensitivity}).
TCR is therefore independent of~$n_c$ within the tested range,
meaning the \emph{absolute number} of affected requests scales
linearly with batch size.

\begin{table*}[t]
  \centering
  \caption{Bit-position sensitivity summary for Qwen3-8B
    (150 trials per row; all metrics are per-row means).
    BS = BERTScore (F1); RL = ROUGE-L (F1).
    None = no effect; Part.\ = partial divergence;
    Comp.\ = complete divergence; Coll.\ = collapse.
    $p_{\text{raw}}$: uncorrected Kruskal--Wallis $p$-value
    (TCR concurrency independence).}
  \label{tab:bit-sensitivity}
  \small
  \begin{tabular}{@{}clrrrrr rrrr @{\hspace{1em}}r@{}}
    \toprule
    Bit & Field &
    \multicolumn{1}{c}{TCR} &
    \multicolumn{1}{c}{TDR} &
    \multicolumn{1}{c}{BS} &
    \multicolumn{1}{c}{RL} &
    \multicolumn{1}{c}{OCR} &
    \multicolumn{1}{c}{None} &
    \multicolumn{1}{c}{Part.} &
    \multicolumn{1}{c}{Comp.} &
    \multicolumn{1}{c}{Coll.} &
    \multicolumn{1}{c}{$p_{\text{raw}}$} \\
    \midrule
    0  & $M_0$ & 19.8\% & 10.5\% & .990 & .940 & 64.0\% &  54 &  93 &  3 &   0 & .852 \\
    1  & $M_1$ & 27.1\% & 14.8\% & .985 & .916 & 69.3\% &  46 &  99 &  5 &   0 & .439 \\
    2  & $M_2$ & 32.4\% & 17.3\% & .983 & .906 & 80.0\% &  30 & 119 &  1 &   0 & .298 \\
    3  & $M_3$ & 39.0\% & 20.9\% & .978 & .880 & 88.0\% &  18 & 128 &  4 &   0 & .120 \\
    4  & $M_4$ & 37.4\% & 21.1\% & .977 & .874 & 81.3\% &  28 & 117 &  5 &   0 & .042 \\
    5  & $M_5$ & 42.0\% & 24.7\% & .973 & .861 & 87.3\% &  19 & 121 & 10 &   0 & .137 \\
    6  & $M_6$ & 42.7\% & 24.8\% & .974 & .864 & 88.0\% &  18 & 123 &  9 &   0 & .055 \\
    7  & $E_0$ & 45.0\% & 25.4\% & .974 & .860 & 92.0\% &  12 & 127 & 11 &   0 & .087 \\
    8  & $E_1$ & 40.9\% & 23.5\% & .975 & .864 & 86.7\% &  20 & 125 &  5 &   0 & .269 \\
    9  & $E_2$ & 44.9\% & 25.9\% & .972 & .850 & 89.3\% &  16 & 129 &  5 &   0 & .034 \\
   10  & $E_3$ & 47.6\% & 28.3\% & .970 & .843 & 91.3\% &  13 & 127 & 10 &   0 & .159 \\
   11  & $E_4$ & 49.9\% & 31.4\% & .960 & .804 & 91.3\% &  13 & 120 & 16 &   1 & .417 \\
   12  & $E_5$ & 56.9\% & 39.6\% & .930 & .713 & 92.7\% &  11 & 111 &  4 &  24 & .793 \\
   13  & $E_6$ & 54.2\% & 38.6\% & .930 & .710 & 92.7\% &  11 & 111 &  4 &  24 & .212 \\
   14  & $E_7$ & 91.9\% & 88.9\% & .749 & .129 & 100.0\% &   0 &  21 &  2 & 127 & .098 \\
   15  & $S$   & 45.9\% & 27.2\% & .971 & .843 & 89.3\% &  16 & 126 &  8 &   0 & .084 \\
    \midrule
    \multicolumn{7}{@{}l}{Total (2{,}400)} &
    325 & 1{,}797 & 102 & 176 & \\
    \bottomrule
  \end{tabular}
\end{table*}

\subsection{Selective Propagation}
\label{sec:eval-selective}

The sensitivity analysis above establishes that, depending on bit
position, a single bit flip alters 20--92\% of requests sharing the
corrupted prefix blocks.
To confirm that corruption does not propagate beyond those blocks,
we run the selective propagation experiment:
for each of the four representative bit
positions, 30 trials inject one bit flip into the first group's
prefix blocks and measure output divergence for both groups.

Across all 120 trials, every request referencing the corrupted
prefix blocks exhibited the expected bit-position-dependent
divergence (affected-group TCR: 28.3\% at bit~0, 43.3\% at bit~6,
90.0\% at bit~14, 46.7\% at bit~15), while requests referencing the
separate prefix blocks produced outputs identical to their clean
baselines
($\mathrm{TCR} = 0.0$ in all 120 trials).  This confirms that
corruption propagates exclusively through the shared block and does
not affect co-located workloads with different prefixes.

\subsection{Temporal Persistence}
\label{sec:eval-persistence}

The preceding analyses characterize the immediate impact on one batch
of concurrent requests.  In practice, a corrupted prefix block remains
cached and is read by every subsequent request sharing the prefix;
this subsection quantifies how long the effect persists.

For each of the four representative bit positions ($p \in \{0, 6, 14,
15\}$) and each of the two models (Qwen3-8B and
DeepSeek-R1), we execute 30 independent runs.
Each run injects one bit flip into the prefix block and then sends 100
sequential requests through the corrupted cache, yielding 240 runs and
24{,}000 post-injection requests in total.  Block survival is monitored at
25-request intervals.  A separate baseline condition (no injection)
is run once per model.

We define a per-request corruption indicator for the $i$-th
request in run~$r$:
\begin{equation}
  \label{eq:corruption-indicator}
  c_i^{(r)} = \mathbb{1}\!\bigl[\hat{y}_i^{(r)} \neq y_i\bigr],
\end{equation}
and report two derived metrics: the \emph{per-request corruption
rate} $\bar{c}_i = R^{-1}\sum_r c_i^{(r)}$, averaged across
$R{=}30$ runs, and the \emph{cumulative affected count}
$C_N^{(r)} = \sum_{i=1}^{N} c_i^{(r)}$.
Unlike TCR (the fraction of concurrent requests affected in one
batch), $\bar{c}_i$ is a binary indicator of whether the $i$-th
sequential request diverged from its baseline.

\subsubsection{No Temporal Decay}
\begin{figure*}[t]
  \centering
  \includegraphics[width=\textwidth]{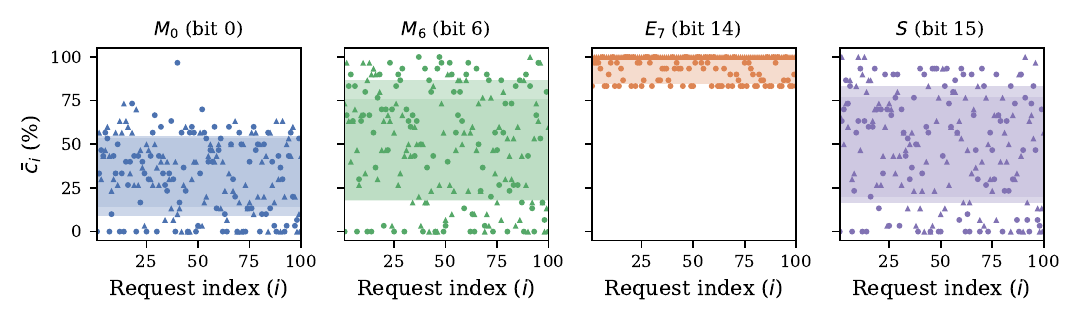}
  \caption{Per-request corruption rate $\bar{c}_i$ over 100 sequential
    requests, split by bit position (circles: Qwen3-8B;
    triangles: DeepSeek-R1).  Shaded bands: mean $\pm 1$ s.d. (darker: Qwen3-8B; lighter: DeepSeek-R1).}
  \label{fig:temporal-persistence}
\end{figure*}

\Cref{fig:temporal-persistence} plots the per-request corruption rate
$\bar{c}_i$ over 100 sequential requests for each bit position and
model.  Across all eight conditions, the corruption rate fluctuates
around a constant mean with no systematic trend.
To test for a monotonic trend without assuming normality, we compute
the Spearman rank correlation between $\bar{c}_i$ and request
index~$i$.  One condition (bit~14 on DeepSeek-R1) is a constant 100\%
rate, for which $\rho$ is undefined; the other seven show weak
negative correlations ($|\rho| \leq 0.24$), none significant after
Bonferroni correction ($p > 0.05$; \cref{tab:temporal-results}).
Over the 100-request horizon, there is thus no intrinsic temporal
decay: the corrupted block affects subsequent requests at a
statistically constant rate.  Whether this extends beyond 100 requests
is left to future work.

\subsubsection{Linear Cumulative Damage}
\begin{figure*}[t]
  \centering
  \includegraphics[width=0.72\textwidth]{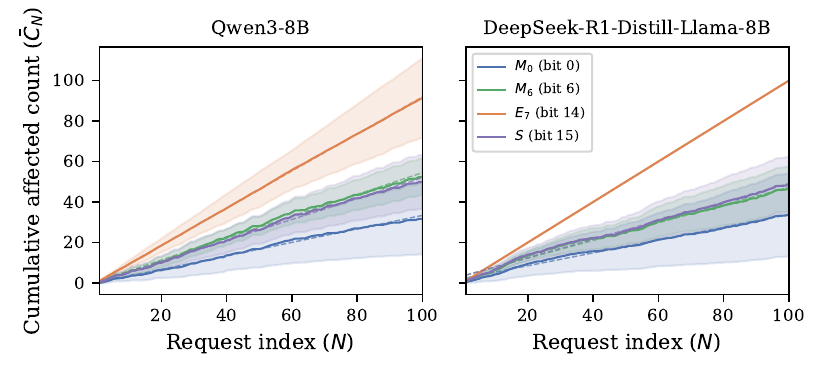}
  \caption{Mean cumulative affected count $\bar{C}_N$ over 100
    requests.  Shaded regions: $\pm 1$ s.d.; dashed lines: OLS linear fits.}
  \label{fig:cumulative-damage}
\end{figure*}

\Cref{fig:cumulative-damage} plots the mean cumulative affected count
$\bar{C}_N$ as a function of~$N$.  All conditions exhibit a linear growth pattern:
ordinary least-squares regression yields
$R^2 \geq 0.992$ for every condition, with the slope approximately equal to the
mean per-request corruption rate.

This linear growth is enabled by sustained residency: across all 240
runs, the corrupted block remained cached at every checkpoint
(25, 50, 75, 100 requests), with no eviction observed---consistent
with vLLM's eviction policy.  In the most severe case (bit~14 on
DeepSeek-R1), the per-request corruption rate reaches 100.0\%: every
subsequent request from the corrupted block is altered.

\begin{table}[t]
  \centering
  \caption{Temporal persistence results per condition.
    \emph{Init~TCR}: initial batch TCR ($n_c{=}2$);
    \emph{CR}: mean per-request corruption rate over 100 requests;
    \emph{Spearman}: rank correlation $\rho$ between $\bar{c}_i$ and
    request index (uncorrected $p$ in parentheses).}
  \label{tab:temporal-results}
  \small
  \resizebox{\columnwidth}{!}{%
  \begin{tabular}{@{}ll@{\hspace{1em}}ccl@{}}
    \toprule
    Model & Bit &
    \multicolumn{1}{c}{Init TCR} &
    \multicolumn{1}{c}{CR} &
    \multicolumn{1}{c}{Spearman $\rho$ ($p$)} \\
    \midrule
    Qwen3-8B  & $M_0$ & .250 & .317 & $-.093$ (.357) \\
    Qwen3-8B  & $M_6$ & .500 & .523 & $-.114$ (.259) \\
    Qwen3-8B  & $E_7$ & .883 & .914 & $-.203$ (.043) \\
    Qwen3-8B  & $S$   & .367 & .499 & $-.112$ (.268) \\
    DeepSeek-R1 & $M_0$ & .167 & .338 & $-.187$ (.063) \\
    DeepSeek-R1 & $M_6$ & .483 & .467 & $-.239$ (.017) \\
    DeepSeek-R1 & $E_7$ & 1.000 & 1.000 & --- \\
    DeepSeek-R1 & $S$   & .350 & .487 & $-.227$ (.023) \\
    \bottomrule
  \end{tabular}}
\end{table}

\subsubsection{Cross-model Consistency}
\Cref{tab:temporal-results} shows that the sensitivity hierarchy
($M_0 < M_6 \approx S < E_7$) is preserved on both models.
Absolute corruption rates differ, but the qualitative
pattern---collapse concentrated at bit~14 and strict temporal
linearity---is consistent across both, supporting the hypothesis
that the hierarchy is a consequence of BF16 encoding rather than
an artifact of a single architecture.
Generalization beyond the 8B scale is discussed in
\cref{sec:discussion-limitations}.

\subsection{Interpretation of Results}
\label{sec:eval-interpretation}

The three properties form a causal chain.  Silent divergence bypasses
content-level quality filters (\cref{sec:eval-sensitivity}) and
selective propagation dilutes aggregate anomaly signals
(\cref{sec:eval-selective}), together constituting \emph{detection
evasion}: an operator monitoring overall service quality sees most
requests unaffected, while the corrupted subset appears individually
legitimate.  Because undetected corruption cannot be remediated,
persistent accumulation (\cref{sec:eval-persistence}) proceeds
unchecked, yielding \emph{damage amplification} bounded only by the
block's residency.
This chain has no analogue in weight corruption, where every inference
is affected uniformly and cross-request comparison readily reveals the
fault~\cite{lin2025gpuhammer}.

\section{\uppercase{Countermeasures}}
\label{sec:countermeasures}

The linear damage growth established in \cref{sec:eval-persistence}
stems from the absence of integrity verification on cached blocks.
Adding integrity verification---detecting and invalidating corrupted
blocks before reuse---directly addresses this gap.
Complementary operator-level mitigations are discussed in
\cref{sec:discussion-implications}.

We impose three requirements on the detection mechanism:
\textbf{complete detection} of any single-bit modification regardless
of position, \textbf{zero false positives} under normal operation, and
\textbf{negligible throughput overhead}.
To the best of our knowledge, no prior work has proposed runtime
integrity verification of cached KV tensors.

\subsection{Detection Mechanism}
\label{sec:countermeasures-design}

\subsubsection{Mechanism}
To meet these requirements, the mechanism verifies block integrity
via hash comparison at two lifecycle events in vLLM's block pool.
\textbf{On cache:} when a prefix block becomes fully computed, a hash
digest is computed over its KV-cache tensor data (copied to CPU) across
all model layers and stored with the block's metadata.  Because block
metadata is registered in the hash map before the GPU tensor write
completes, digest computation is deferred until after
\texttt{execute\_model()} returns.
\textbf{On access:} every cache hit recomputes and compares the digest;
a mismatch evicts the block and forces recomputation from clean model
weights (a single prefill pass).

The mechanism is activated by an optional flag and comprises
approximately 100~lines of new code across five source files in
vLLM v1.
We adopt SHA-256 (already present in vLLM's dependency chain) as
the digest algorithm.  Because bit flips alter only a small number
of bits rather than arbitrary bytes, cryptographic collision
resistance is unnecessary; faster alternatives (xxHash3, BLAKE3) are
drop-in replacements.
The end-to-end overhead is quantified in \cref{sec:countermeasures-overhead}.

\subsubsection{Security Guarantee}
The mechanism guarantees that any single-bit corruption present at the
start of a scheduling cycle is detected and invalidated before the
block is served in that cycle.
For $N$ requests served from a corrupted block during its residency,
detection bounds cumulative affected requests by one batch,
independent of~$N$, reducing $\Theta(N)$ growth to $O(1)$.
Residual gaps---time-of-check-to-time-of-use (TOCTOU) windows,
deferred sealing, and
GPU floating-point non-determinism after recomputation---are discussed
in \cref{sec:discussion-limitations}.

\subsection{Detection Evaluation}
\label{sec:countermeasures-detection}

To validate implementation soundness---correct serialization,
block-boundary alignment, and digest storage/retrieval---we run
120 injection trials on Qwen3-8B ($n_c{=}2$; same four bit
positions and 30 trials per position as in \cref{sec:eval-persistence}).
All 120 injected bit flips were detected and invalidated before
block reuse, with zero false positives.

\subsection{Throughput Overhead}
\label{sec:countermeasures-overhead}

We measure throughput overhead by comparing generation speed with and
without the integrity mechanism enabled.  Each condition consists of
30~runs of 100~sequential requests (batch size~1) on Qwen3-8B (BF16)
with no injection, yielding
3{,}000~data points per condition.  The metric is tokens per second
per request, measured on follow-up requests only (warm-up, baseline,
and injection phases excluded).

Mean throughput was $13.80 \pm 0.07$\,tok/s with the mechanism
enabled and $13.69 \pm 0.05$\,tok/s with it disabled.
The 0.11\,tok/s difference is comparable to each condition's
standard deviation and attributable to measurement noise, so
the overhead of integrity verification is negligible under
sequential serving.
Across all 3{,}000 verified requests with the mechanism enabled, zero
false positives were observed.
Measurement caveats are discussed in \cref{sec:discussion-limitations}.

\section{\uppercase{Discussion}}
\label{sec:discussion}

The threat chain identified in \cref{sec:eval-interpretation}---detection
evasion enabling unchecked damage accumulation---motivates both
operational mitigations and further investigation of the remaining
gap between our SFI analysis and physical exploitation.

\subsection{Practical Implications}
\label{sec:discussion-implications}

Existing operational mitigations address adjacent layers: hardware ECC
(\cref{sec:bg-rowhammer}) mitigates bit-flip risks in KV-cache memory,
and vLLM's \texttt{cache\_salt}~\cite{vllm_pr17045}---introduced
against timing side channels~\cite{song2025earlybirdleak}---eliminates
cross-salt-group sharing but forfeits the caching benefit.  Our
checksum is orthogonal, preserving sharing while detecting corruption.
Our results motivate three additional measures.
\textbf{Output-quality monitoring} (repetition ratio or perplexity
thresholds) detects collapse-mode corruption (bit~14) but is
\emph{inherently blind to silent divergence}: 13 of 16 bit positions
produce coherent outputs that evade content-level quality
filters (\cref{sec:eval-sensitivity}).
\textbf{TTL-based block recomputation} caps cumulative damage at
$O(N_{\mathrm{TTL}})$ without integrity-checking infrastructure, at the
cost of increased prefill overhead.
\textbf{Runtime integrity verification}---our checksum-based
countermeasure (\cref{sec:countermeasures-design})---is the only
measure that addresses silent divergence directly without sacrificing
cache effectiveness, reducing cumulative damage to $O(1)$.

\subsection{Limitations}
\label{sec:discussion-limitations}

\subsubsection{Gap from SFI to Physical Exploitation}
Our ideal-bit-targeting assumption (\cref{sec:threat-model})
establishes an upper bound on impact severity by deliberately
relaxing physical Rowhammer constraints---row adjacency, fixed flip
direction, and undocumented address
mappings~\cite{kim2014rowhammer,lin2025gpuhammer}.
End-to-end exploitation requires three capabilities:
(1)~\textbf{cache-presence detection}, demonstrated via side-channel
attacks~\cite{song2025earlybirdleak,wu2025promptpeek,pennas2026cachesolidarity};
(2)~\textbf{physical targeting} of a specific prefix
block---partially addressed by
\textsc{CacheTrap}~\cite{alnahian2025cachetrap} on transient
caches but not long-lived prefix blocks;
(3)~\textbf{bit flip}, shown feasible by \textsc{CacheTrap} on KV-cache
value tensors via \textsc{GPUHammer}~\cite{lin2025gpuhammer}.

\subsubsection{Generality}
We evaluate two 8B-parameter BF16 models on a single GPU, targeting
value tensors.  Our basis for generality is mechanistic: the per-bit
sensitivity hierarchy follows from the BF16 encoding
(\cref{sec:bg-bf16}) and should hold across models, though
quantitative rates may vary; the agreement between Qwen3-8B and
DeepSeek-R1---models from different families
(\cref{sec:eval-persistence})---corroborates this.  Generalization to
other architectures (e.g., Mistral), larger models (70B+),
lower-precision formats (e.g., FP8, INT4), multi-bit flips, multi-GPU
parallelism, and diverse tasks---including safety-aligned system
prompts, whose KV blocks are shared in production---remains
unexplored.  Key-side flips may yield qualitatively different failure
modes (e.g., attention misrouting).  The same hierarchy also applies
to non-adversarial faults (DRAM retention or soft errors), with
implications for reliability engineering.

\subsubsection{Residual Detection Gaps}
The checksum verifies blocks at CPU-side scheduling time, leaving a
one-cycle window: a flip after verification but before the GPU-side
attention read, or a cache hit on a newly registered block before
deferred sealing.  A GPU-side hashing kernel would narrow both gaps.
Separately, block recomputation after checksum-triggered eviction may
introduce bit-level differences due to non-deterministic GPU
floating-point reduction; injection-free controls (210 runs) confirmed
this is inherent non-determinism, not residual corruption.

\subsubsection{Measurement Caveats}
Throughput overhead (\cref{sec:countermeasures-overhead}) is measured
at batch size~1 on unified memory (NVIDIA DGX Spark).
On discrete GPUs, each verification of the ${\approx}2.25$\,MiB
per-block data (Qwen3-8B) requires a PCIe round trip, and concurrent
cache hits would multiply the cost, making the reported overhead a
lower bound.

\subsection{Ethical Considerations}
\label{sec:ethics}

All experiments used locally hosted models with software fault
injection; no real user traffic or production systems were affected.
Prior to publication, we disclosed the vulnerability to the vLLM
maintainers (April 8, 2026) via a GitHub Security Advisory with a
proposed patch; they declined it, regarding the threat scenario as not
plausible in current deployments.
Source code and data will be released upon publication, excluding
tools for physical Rowhammer exploitation or GPU memory address
translation.

\section{\uppercase{Related Work}}
\label{sec:related-work}

\begin{table*}[t]
  \centering
  \caption{Positioning relative to representative bit-flip and
    KV-cache attacks.  SD, SP, and PA denote whether a work
    characterizes the silent divergence, selective propagation, and
    persistent accumulation we identify; Def.\ denotes whether
    an integrity defense is proposed.
    \checkmark~= addressed; --~= not addressed;
    n/a~= not applicable (different target surface).}
  \label{tab:comparison}
  \small
  \setlength{\tabcolsep}{4.5pt}
  \begin{tabular}{@{}lllcccc@{}}
    \toprule
    Work & Target & Fault model & SD & SP & PA & Def. \\
    \midrule
    \textsc{BFA}~\cite{rakin2019bfa} & DNN weights & SFI (multi-bit) & n/a & n/a & n/a & -- \\
    \textsc{PrisonBreak}~\cite{coalson2025prisonbreak} & LLM weights & Rowhammer, GPU (multi-bit) & n/a & n/a & n/a & -- \\
    \textsc{CacheTrap}~\cite{alnahian2025cachetrap} & Transient KV & Rowhammer, GPU (1-bit) & -- & -- & -- & -- \\
    \textsc{Kelle}~\cite{xia2025kelle} & KV (eDRAM) & Retention (non-adversarial) & coarse & -- & -- & \checkmark \\
    \textsc{History Swapping}~\cite{ganesh2025kvcachemanipulation} & KV blocks & Block read/write swap & -- & -- & -- & -- \\
    \textsc{CacheMe}~\cite{wu2026cacheme} & Cache index & API-level & n/a & n/a & n/a & n/a \\
    \midrule
    \textbf{This work} & Shared prefix KV & SFI (1-bit) & \checkmark & \checkmark & \checkmark & \checkmark \\
    \bottomrule
  \end{tabular}
\end{table*}

\subsection{Bit-Flip Attacks}
Rowhammer (\cref{sec:bg-rowhammer}) has been exploited to
corrupt a range of in-memory targets:
page tables for privilege escalation~\cite{seaborn2015rowhammer},
browser-based remote exploitation~\cite{gruss2016rowhammer},
cryptographic keys via read-side leakage~\cite{kwong2020rambleed},
bypasses of in-DRAM mitigations~\cite{frigo2020trrespass},
nested pointers for arbitrary data leakage~\cite{tobah2024gadgethammer},
and---on GPUs---GDDR6 memory~\cite{lin2025gpuhammer} and page tables
for cross-context privilege escalation~\cite{lin2026gpubreach}.

Neural network weights are another instance of in-memory data, and
bit-flip attacks on them have followed two
complementary tracks.
SFI studies characterize vulnerability under ideal bit targeting:
\textsc{BFA}~\cite{rakin2019bfa} established the methodology for DNN
weights, and subsequent work applied it to LLMs, collapsing accuracy
via evolutionary search~\cite{das2025genbfa} or a single stealthy
flip~\cite{guo2025sbfa}.
Physical Rowhammer studies demonstrate end-to-end exploitation:
\textsc{DeepHammer}~\cite{yao2020deephammer} attacked DNN weights on
CPU DRAM, \textsc{OneFlip}~\cite{li2025oneflip} backdoored
full-precision DNNs with a single flip, and
\textsc{PrisonBreak}~\cite{coalson2025prisonbreak} extended this to GPU
GDDR6, jailbreaking safety-aligned LLMs; further demonstrations include
remote bit search~\cite{yan2025bitsifter} and targeted content
steering~\cite{guo2026tfl}.

All of these studies, whether SFI or physical, target
\emph{static model weights}, whose corruption affects every
subsequent inference uniformly; none considers the
\emph{dynamic, shared} KV-cache structures that our work
addresses, where corruption impacts only the requests sharing
the corrupted block.

\subsection{KV-Cache Security and Resilience}
Two studies target KV-cache tensor data through bit flips.
\textsc{CacheTrap}~\cite{alnahian2025cachetrap}, the closest to our
work, shares a value-side single-bit fault model but targets a single
request's transient cache rather than shared prefix blocks, and
evaluates classification tasks where corruption is binary rather than
coherent but altered natural language.
\textsc{Kelle}~\cite{xia2025kelle} observes a coarse-grained
MSB\,$>$\,LSB sensitivity hierarchy for KV-cache bit flips, but under
non-adversarial eDRAM retention errors in an edge setting, without
per-bit-position characterization or shared prefix blocks.

The remaining studies target different attack surfaces.
\textsc{History Swapping}~\cite{ganesh2025kvcachemanipulation}
corrupts KV-cache tensor data but assumes direct read--write access to
swap entire blocks, not single-bit corruption via hardware faults.
\textsc{CacheMe}~\cite{wu2026cacheme} targets cache indexing through
API-level attack vectors rather than the cached tensor data itself.

None of the four works considers silent divergence, selective
propagation, or persistent accumulation---the properties central
to our threat characterization.
A direct empirical comparison is not feasible---these works target
disjoint surfaces, with no shared benchmark or victim model---so
\cref{tab:comparison} positions our work qualitatively.

\section{\uppercase{Conclusion}}
\label{sec:conclusion}

This paper presents the first systematic severity characterization of
shared prefix blocks in LLM serving systems under adversarial
single-bit corruption.
Across 2{,}400 fault injection trials and 24{,}000 post-injection
requests on two 8B-parameter models, a single bit flip in a shared
prefix block yields three compounding properties---silent divergence,
selective propagation, and persistent accumulation---that together
enable detection evasion and damage amplification bounded only by
block residency, with no analogue in weight corruption.
Our checksum-based countermeasure breaks this chain, detecting all
120 injected faults with zero false positives and negligible overhead,
bounding cumulative damage to one batch regardless of residency.
Prefix blocks therefore warrant integrity protection before the
SFI-to-physical gap is closed.

Two directions merit further investigation.  First, the severity
characterization should be broadened to key-side sensitivity,
lower-precision formats, and multi-GPU tensor parallelism.  Second,
the reachability gap remains open: analyzing block placement
predictability and combining GPU virtual-to-physical address
translation with demonstrated cache detection techniques would enable
assessment of end-to-end exploitation feasibility.

\ifblind\else
\section*{\uppercase{Acknowledgements}}
This work was supported in part by Palo Alto Networks.
The manuscript was drafted and edited with the assistance of
Claude Opus 4.6~\cite{anthropic2026opus46}, a large language model
developed by Anthropic.  The authors are responsible for the scientific content,
experimental design, and all claims.
\fi

\bibliographystyle{scitepress/apalike}
{\small
\bibliography{references}}

\end{document}